\documentstyle[aaspp4,12pt,psfig]{article}
\lefthead{Bower et al.}
\righthead{Linear Polarization of Sgr~A*}
\begin{document}


\newcommand\degd{\ifmmode^{\circ}\!\!\!.\,\else$^{\circ}\!\!\!.\,$\fi}
\newcommand{\etal}{{\it et al.\ }}
\newcommand{\uv}{(u,v)}
\newcommand{\rdm}{{\rm\ rad\ m^{-2}}}

\title{The Linear Polarization of Sagittarius A* II.  \\
VLA and BIMA Polarimetry at 22, 43 and 86 GHz}
\author{Geoffrey C. Bower\altaffilmark{1,2}, 
Melvyn C.H. Wright\altaffilmark{3},
Donald C. Backer\altaffilmark{3}, 
\& Heino Falcke\altaffilmark{2}}
\altaffiltext{1}{National Radio Astronomy Observatory, P.O. Box O, 1003 
Lopezville, Socorro, NM 87801}
\altaffiltext{2}{Max Planck Institut f\"{u}r Radioastronomie, Auf dem 
H\"{u}gel 69, D 53121 Bonn Germany}
\altaffiltext{3}{Astronomy Department \& Radio Astronomy Laboratory, 
University of California, Berkeley, CA 94720}



\begin{abstract}

We present a search for linear polarization at 22 GHz, 43 GHz and 86
GHz from the nearest super massive black hole candidate, Sagittarius A*.  
We find upper limits to the
linear polarization of 0.2\%, 0.4\% and 1\%, respectively.  These
results strongly support the conclusion of our centimeter wavelength
spectro-polarimetry that Sgr A* is not depolarized by the interstellar
medium but is in fact intrinsically depolarized.

\end{abstract}


\section{Introduction}

The compact non-thermal radio source Sgr~A* is recognized as one of the
most convincing massive black hole candidates (Maoz \markcite{maoz98}
1998).  Recent results from stellar proper motion studies indicate that
there is a dark mass of $\sim 2.6 \times 10^6 M_{\sun}$ enclosed within
0.01 pc (Genzel \etal \markcite{genze97} 1997, Ghez \etal
\markcite{ghez98} 1998).  Very long baseline interferometry studies at
millimeter wavelengths have shown that the intrinsic radio source
coincident with the dark mass has a size that is less than 1 AU and a
brightness temperature greater than $10^9$ K (Rogers \etal
\markcite{roger94} 1994, Bower \& Backer \markcite{bower98} 1998, Lo
\etal \markcite{lo98} 1998, Krichbaum \markcite{krich98} \etal 1998).  
Together these points are compelling
evidence that Sgr~A* is a cyclo-synchrotron emitting region surrounding
a massive black hole.  Nevertheless, specific details of the excitation
of high energy electrons, their distribution and the accretion of
infalling matter onto Sgr~A* are unknown (e.g., Falcke, Mannheim \&
Biermann \markcite{falck93} 1993, Melia \markcite{melia94} 1994,
Narayan \etal \markcite{naray98} 1998, Mahadevan \markcite{mahad98}
1998).

We have recently demonstrated that Sgr A* is not linearly polarized at
a level of 0.2\% at 4.8 and 8.4 GHz (Bower \etal \markcite{bower99a}
1999, hereafter Paper~I).  This spectro-polarimetric result excludes
rotation measures up to $10^7 \rdm$.  Interstellar depolarization
in the scattering region (Frail \etal \markcite{frail94} 1994,
Yusef-Zadeh \etal \markcite{yusef94} 1994, Lazio \& Cordes
\markcite{lazio98} 1998) is unlikely but not completely excluded by
these observations.  Interstellar depolarization can occur if the scale
of turbulent fluctuations in the scattering medium are on the order of
$10^{-4} {\rm \ pc}$.  Although this scale is probably too large, it is
not fully excluded by observations.  The millimeter polarimetry that we
describe in this paper directly addresses the significance of
interstellar depolarization on these scales.

Our recent detection of circular polarization in Sgr A* gives
particular relevance to the question of the level of intrinsic
polarization (Bower, Falcke \& Backer \markcite{bower99b} 1999).
Typically, AGN display integrated circular polarization that is an order of
magnitude or more less than the integrated linear polarization 
(Weiler \& de Pater \markcite{weile83} 1983).
This is not only the consequence of beam dilution.  
In the case of the VLBI detection of circular polarization
in a compact knot in 3C 279, the circular polarization is less
than the co-spatial linear polarization by a factor of $\sim 10$ (Wardle
\etal \markcite{wardl98} 1998).  That is, there are no known regions
in jets with high circular polarization and low linear polarization.
Therefore, the presence of a large circular to
linear polarization ratio in Sgr A* is an unsolved and
intriguing radiative transfer
problem.  We discuss later some of the models that may account for
this ratio.

In \S 2 we present VLA\footnote{The VLA is an instrument of the
National Radio Astronomy Observatory.  The NRAO is a facility of the
National Science Foundation, operated under cooperative agreement with
Associated Universities, Inc.}
and  BIMA\footnote{The BIMA
array is operated by the Berkeley-Illinois-Maryland Association under
funding from the National Science Foundation} array polarimetry.  There is no
detected polarization for Sgr A* at 22, 43 and 86 GHz.  In \S 3 we
demonstrate that interstellar depolarization at these frequencies is
extremely unlikely.  We consider the consequences of an intrinsically
unpolarized Sgr A* in \S 4.

\section{Observations and Data Reduction}

\subsection{VLA Observations at 22 GHz and 43 GHz}

We observed Sgr A* on 3 February 1997 at 22 GHz and 43
GHz using the VLA
The array was in the BnA configuration.  Data were obtained in two
50 MHz wide intermediate frequency (IF) 
bands at 22.435 and 22.485 GHz, and 43.315 and 43.365 GHz, respectively.
The 27-element array was divided into two sub-arrays that observed
simultaneously at 22 GHz and 43 GHz.
The flux density scale was set by assuming standard flux densities for 
3C 286.
Hourly observations of B1730-130 were
used to measure antenna-based gain amplitude fluctuations and to
determine the antenna-based polarization leakage terms, following
standard practices.  Absolute position angle calibration was not
possible due to errors in the cross-correlation data for 3C 286.  
All measured position angles were rotated so that the position angle
for B1730-130 was set to 0.

Sgr A* and the compact source B1741-312 were each
observed twice an hour for 7 hours.  The compact source B1921-293 was
observed at 43 GHz once an hour for 4 hours.  Total and polarized
intensities in each IF band were measured as the best-fit Gaussian
in the $I$ and $P$ images (Table~\ref{tab:vlapol}).  The quoted
errors are rms errors from the fit.  We also report the off-source
maximum value in the polarized image, $P_{lim}$ in flux units and
$p_{lim}$ as a fraction of the total intensity.  A real detection must
be more than twice this value to be believable.

The measured polarizations for Sgr A* are many times the rms image
noise, which is on the order of 0.2 mJy.  
However, there is a significant contribution from multiplicative
errors.  These errors principally derive from variations in the 
polarization leakage terms (Holdaway, Carilli \& Owen \markcite{holda92} 1992).
The effect of the $D$-term errors is to scatter a fraction of the total 
intensity into the polarized intensity map.  Typically, at 
centimeter wavelengths the VLA can achieve a fractional error of
$\sim 0.1\%$ (e.g., Bower\etal 1999a).
The smaller number of antennas and poorer performance of the array
at 22 GHz and 43 GHz will lead to larger fractional polarization 
errors.

Comparing
results between IF bands is not a reliable method for determining
fractional errors.  The dominant sources of $D$-term errors are common
to both antennas.  Hence, we see variations between IFs for bright
sources that are fully consistent with the thermal noise.

Two factors indicate that the measured polarization for Sgr~A*
is an upper limit rather than a detection.
We show in Figure~1 a 43 GHz image of Sgr~A* with polarization vectors
overlaid.  First, 
there is large variation in the polarization position angles over
the source.  This is also true in the 22 GHz images.
Second, the sidelobes and noise peaks are polarized at a level comparable to
the central source.
Off-source peaks in the $P$ maps are as large as 
the measured polarization.  This implies
fractional polarization errors of 0.2\% and 0.4\% at 22 GHz and 43 GHz,
respectively.

\subsection{BIMA Observations at 86 and 90 GHz}

Polarimetric observations of Sgr A* were obtained with the BIMA array
(Welch \etal \markcite{welch96} 1996) on three dates, 10 March
1998, 14 March 1998 and 19 December 1998.  The array was in the A
configuration producing  projected baselines for Sgr A* in the range 20
to 520 $k\lambda$.  Continuum bandwidths were 800 MHz in lower and
upper IF sidebands centered at 86.582 GHz and 90.028 GHz.  Standard
antenna amplitude gains were applied.

Each receiver is sensitive to linear polarization.  Quarter-wave plates
were installed on all antennas such that the receivers can be switched
between linear, right circular (RCP) or left circular (LCP)
polarization.  One antenna observed linear polarization
continuously, while the other antennas were switched between RCP and
LCP using a Walsh-function pattern to optimize the visibility
coverage in parallel- and cross-hand correlations (Wright
\markcite{wrigh95} 1995, Wright \markcite{wrigh96} 1996).  The data
were self-calibrated for both RCP and LCP with respect to the antenna
observing linear polarization.  Because RCP and LCP is detected with
the same receiver in each antenna, there is no phase-offset between
the parallel hand visibilities.  Hence, the 
absolute position angle is correctly determined without any
further calibration.

For all three observations instrumental leakage was calibrated from
observations of strong unresolved sources. The instrumental leakage is
stable to about 0.4\% rms.  This implies that the minimum error in
the polarization maps will be 0.4\%.  If variations in the $D$-terms
are correlated, the error could be over 1\% (Holdaway, Carilli \& Owen 
\markcite{holda92} 1992).
For the March observations, we used
$D$-term solutions from spectral-line observations of the Orion SiO
maser on 28 January 1998 and 25 February 1998 (Rao \etal
\markcite{rao98} 1998).  The average difference per antenna between the
Orion maser $D$-term solutions is 1.3\%, implying a
minimum error in the polarization of $\sim 0.4\%$ if the variations
between antennas are uncorrelated.  The average 
difference between the two Orion maser and calibrator $D$-term 
solutions is similar.  This implies that we are not strongly affected
by variations in the $D$-term solutions over the bandpass.  
Because solutions
were found for a spectral line, they were available only at a single
IF frequency.  For the 19 December 1998 observations, we used solutions found for 3C 273
observed on 21 November 1998 in the C array. These data showed better agreement
between the two IF bands than the solutions found from
interleaved observations of B1730-130.  A similar level of variation in
the $D$-term solutions was found for these observations.

We summarize the total and polarized intensity in
Table~\ref{tab:bimapol}.  The reported errors are estimated from fits
to the corrected parallel- and cross-hand visibility data.  
As is the case with the VLA data,
these are underestimates because they do not
take into account amplitude calibration and polarization leakage term
errors.  
We estimate the total error by the level of off-source peaks in the
polarization maps.  These are on the order of 20 mJy, or 1\%, for Sgr A*.  
This is consistent with the results of Rao \etal \markcite{rao98},
in which the linear polarization limit is 1.5\%.
Therefore,
we consider the measured polarization for Sgr~A* to be an upper limit of 1\%.

In Figure~2 we summarize all upper limits to the polarization of Sgr A*
from Paper~I and from this paper.

\section{Interstellar Depolarization}

A very large rotation measure (RM) will rotate the position angle of 
linear polarization through the observing band.
However, bandwidth depolarization is unlikely to occur in these observations.
The maximum rotation measure detectable in the continuum band of
these experiments is $1.3 \times 10^6 \rdm$, $8.4 \times 10^6 \rdm $
and $4.8 \times 10^6 \rdm$ at 22 GHz, 43 GHz and 86 GHz, respectively.
The spectro-polarimetric observations in Paper~I would have detected a
signal at these RMs if they were present.



We argued in Paper~I that the scattering medium will depolarize the
source if variations in the RM lead to a phase change of $\pi$
radians.  The required RM variations at 22 GHz, 43 GHz and 86 GHz are
$1.8 \times 10^4 \rdm$, $6.4 \times 10^4 \rdm$ and $2.7 \times 10^5
\rdm$.  The known variations in the RM in the Galactic Center region
(Yusef-Zadeh, Wardle \& Parastaran \markcite{yusef97} 1997) are not
sufficient to depolarize Sgr A* at 4.8 GHz and 8.4 GHz (Paper~I).
Therefore, we must only consider whether the depolarization conditions
could arise in the scattering medium around SgrA*.

The angular broadening of images of masers near the Galactic Center and 
Sgr A* is most likely associated with the ionized skins of molecular 
clouds.  The ionization mechanism is either photo-ionization by hot
stars (Yusef-Zadeh \etal \markcite{usef94} 1994) or contact with diffuse,
hot gas (Lazio \& Cordes \markcite{lazio98} 1998).
There are two relevant length scales for the structure of these 
scattering screens: the thickness of the ionized skins, 
$l_{skin}\sim 10^{-4} {\rm\ pc}$ which was derived 
by Yusef-Zadeh \etal \markcite{yusef94} (1994); and the outer scale of the 
turbulent spectrum of electron density fluctuations within these skins, 
$l_0 \sim 10^{-7}{\rm\ pc}$ which was derived by Lazio \& 
Cordes \markcite{lazio98} (1998). The small outer scale in relation to the 
skin depth suggests that these layers may contain many independent turbulent 
cells. The small angular scale of these cells, $l_0/8$ kpc $\sim 0.02$ mas, 
means that they can depolarize a linearly polarized signal owing to their 
random Faraday rotations.  The rms RM along independent lines of 
sight through a single skin  will depend on $l_0\sqrt{l_{skin}/l_0}$. This rms 
will be about some mean if the magnetic field is uniform in the 
skin or about zero if the field is random.  If our line of sight 
traverses $N$ skins, then the equivalent path length for the 
rms RM estimation is $L = \sqrt{N l_{skin} l_0}$. This path length is 
less than $10^{-5} {\rm\ pc}$ for $N < 10$ scattering screens.

The constancy of maser image anisotropy over $\la 10 {\rm\ arcsec}$ angular 
scales suggests that the average perpendicular to the line of sight 
magnetic field imbedded in these skins is uniform over physical scales 
of $\la 1$ pc (Yusef-Zadeh \etal \markcite{yusef99} 1999). This scale is 
a significant fraction 
of the size of molecular clouds in the Galactic Center region. 
Hence, the variations on greater scales may be the result of scattering by 
physically distinct regions.  This uniformity then requires the rms RM to 
be about some mean RM (with contributions from density
alone) rather than about zero (with contributions from density and field).

We show now that for  $L$ as large as $10^{-4} {\rm\ pc}$, depolarization
in the scattering medium and energy equipartition 
between the magnetic field and particle energy require that either or both the
electron density and magnetic field strength exceed the peak values
measured in the Galactic Center region.  These two conditions require
\begin{equation}
n_e=7.3 \times 10^4 {\rm\ cm^{-3}} {\rm\ RM_4}^{2/3} L_{-4}^{-2/3} T_4^{-1/3}
\end{equation}
and 
\begin{equation}
B=1.6 {\rm\ mG} {\rm\ RM_4}^{1/3} L_{-4}^{-1/3} T_4^{1/3},
\end{equation}
where ${\rm RM_4}$ is the rotation measure in units of $10^4 {\rm\ rad\ m^{-2}}$,
$L_{-4}$ is the length scale in units of $10^{-4} {\rm\ pc}$ and 
$T_4$ is the electron temperature in units of $10^4$ K.
Mehringer \etal
\markcite{mehri93} (1993) showed that ionized densities in H~II regions
are significantly less than $10^5 {\rm \ cm^{-3}}$ on arcsecond
scales.  Magnetic field strengths measured with OH masers in dense
molecular regions are on the
order of a few milliGauss (Yusef-Zadeh \etal \markcite{yusef99} 1999).

At 22 GHz and assuming $T_4=1$, we find $B\approx 2 {\rm\ mG}$ and $n_e \approx 10^5 {\rm\ cm^{-3}}$,
which exceeds the observed upper limit on electron density.  At 86 GHz,
$B\approx 5 {\rm \ mG}$ and $n_e \approx 7 \times 10^5
{\rm\ cm^{-3}}$.  For the case of $L\sim10^{-7}{\rm\ pc}$,
depolarization of the 22 GHz radiation requires $B\approx 15 {\rm
\ mG}$ and $n_e \ga 10^7 {\rm\ cm^{-3}}$.   The case is much worse
at 86 GHz.  Increasing the electron
temperature does not allow depolarization:  it 
leads to lower electron densities but higher magnetic fields.  
Therefore, we consider it extremely unlikely that Sgr A* is
depolarized by the interstellar medium.

These electron densities correspond more closely to what we expect from
a sub-parsec accretion flow onto Sgr A* (Melia \markcite{melia94} 1994, 
Melia \& Coker \markcite{melia99} 1999, Quataert, Narayan \& Reid 
\markcite{quata99} 1999).  
As Melia and Coker show, densities in excess of $10^5 {\rm\ cm^{-3}}$
appear at radii less than $\sim 0.01$ pc.  
We demonstrated in Paper~I that this can easily lead to very high
RMs and that depolarization will occur if the accretion region
is sufficiently turbulent.
However, the detailed character of the accretion region is not well-known.
The geometry, volume filling factor and degree of turbulence are
poorly constrained.

\section{An Intrinsically Weakly Polarized Sgr A*}

The degree of linear polarization in AGN typically rises with
frequency.  Aller, Aller \& Hughes \markcite{aller92} (1992) showed
that in their flux-limited sample $\sim 40\%$ of AGN have polarization
fractions less than 1\% at 4.8 GHz while $\sim 10\%$ of the same sample
have polarization fractions less than 1\% at 14.5 GHz.  All sources in
the sample have detected polarization fractions greater than 0.2\% at
14.5 GHz.  This includes 3C 84 which has an average polarization
fraction at 4.8 GHz of $0.03 \pm 0.01 \%$.  
A polarization increase with frequency can be explained
by the high RMs present in some radio cores (Taylor \markcite{taylo98}
1998), the increased prominence of shocked regions and the decreased
synchrotron opacity (Stevens, Robson \& Holland \markcite{steve96}
1996).  We note that a flux-limited
sample of this kind is biased towards powerful, beamed sources which
may have different polarization properties than weaker unbeamed
sources.  The polarization properties of these weaker sources are not
well-studied due to their low flux densities.  There is no
high-frequency polarization study of a volume-limited sample for weaker
sources.  However, Rudnick, Jones \& Fiedler \markcite{rudni86} (1986)
did observe a sample of ``weak'' cores with flat spectra.  They found
that even at 15 GHz many of these sources were unpolarized at a level
of $\sim 1\%$.

The absence of linear polarization in Sgr A* from 4.8 GHz to 86 GHz can
be explained with the presence of thermal electrons or with significant
magnetic field cancellation.  The thermal electrons may be outside the
emission region (in the accretion flow, as discussed above, but not
in the scattering medium) or
may be coincident with the emission region.  This latter case is appealing
because it may be able to account for the presence of circular
polarization through the conversion of linear polarization to circular
polarization (Bower, Falcke \& Backer \markcite{bower99b} 1999,
Pacholczyk \markcite{pacho77} 1977, Jones \& O'Dell
\markcite{jones77} 1977).

Magnetic field cancellation could occur as the result of a tangled
field or a circularly symmetric field orientation.  The former is
typically assumed to depolarize radio jets.  This requires for Sgr~A*
that the emission region consist of $\left(70/0.2\right)^2\approx 10^5$
independent B-field cells.  The latter case may arise if
the emission originates in a quasi-spherical inflow (e.g., an ADAF
model).  Magnetic field cancellation is an unlikely depolarization
mechanism if the circular polarization is intrinsic to the source
(e.g., Wilson \& Weiler \markcite{wilso97} 1997).  However, if the
circular polarization arises from interstellar propagation effects
(Macquart \& Melrose \markcite{macqu99} 1999), then magnetic field
cancellation is a possible explanation.  In this case, the absence of
linear polarization argues against a strong shock origin for the total
flux variability in Sgr A* (Wright \& Backer \markcite{wrigh93} 1993,
Falcke \markcite{falck99} 1999).  Total flux variability in AGN comes
about from the presence of shocks which order the magnetic field and
accelerate particles in the relativistic jet leading to linearly
polarized emission (Marscher \& Gear \markcite{marsc85} 1985).

We have shown here that Sgr A* is not linearly polarized to the current
limits of instrumental sensitivity at 22, 43 and 86 GHz.  The
possibility is remote that Sgr A* is externally depolarized.
However, the linear and circular
polarizations are unique to Sgr A*.  Explaining that relationship may
reveal significant details for the emission region and environment of
Sgr A*.

\acknowledgements 
This work was partially supported by NSF Grant AST-9613998 to the
University of California, Berkeley.
HF is supported by DFG grant Fa 358/1-1\&2.

\newpage

\begin{figure}
\mbox{\psfig{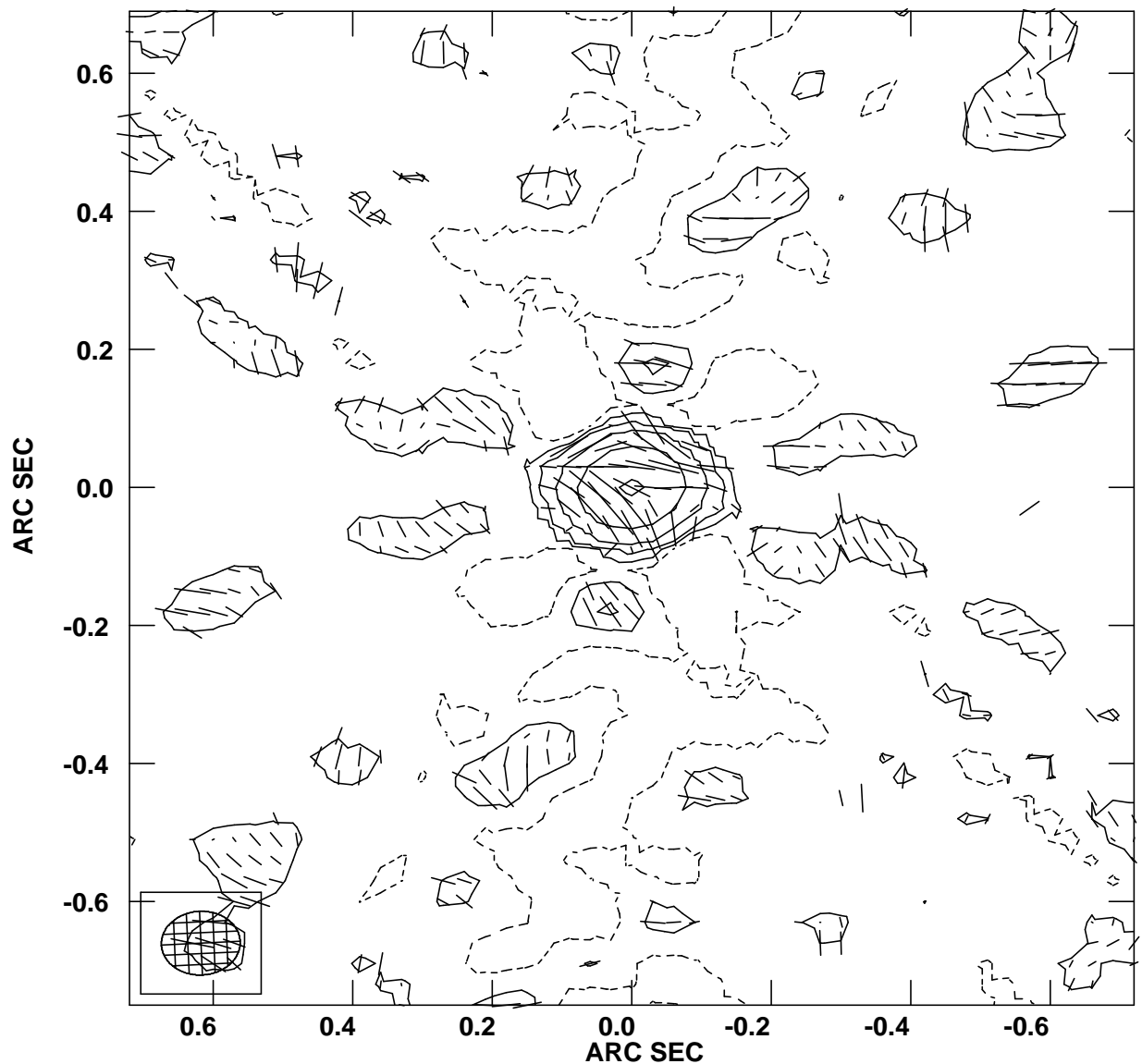}}
\caption{A total intensity image of Sgr A* with polarization
contours overlaid from IF 1 at 43 GHz.  The scatter in the polarization
vectors over the compact source and the strength of the off-source
polarization indicate that the polarization peak is an upper limit.
The total intensity contours are -1\%, 1\%, 3\%, 10\%, 30\% and 90\%
of the peak intensity.  A polarization vector one arcsecond long
represents a polarized intensity of 33.3 mJy/beam.  The synthesized
beam is shown in the lower left corner.}
\end{figure}

\begin{figure}
\mbox{\psfig{figure=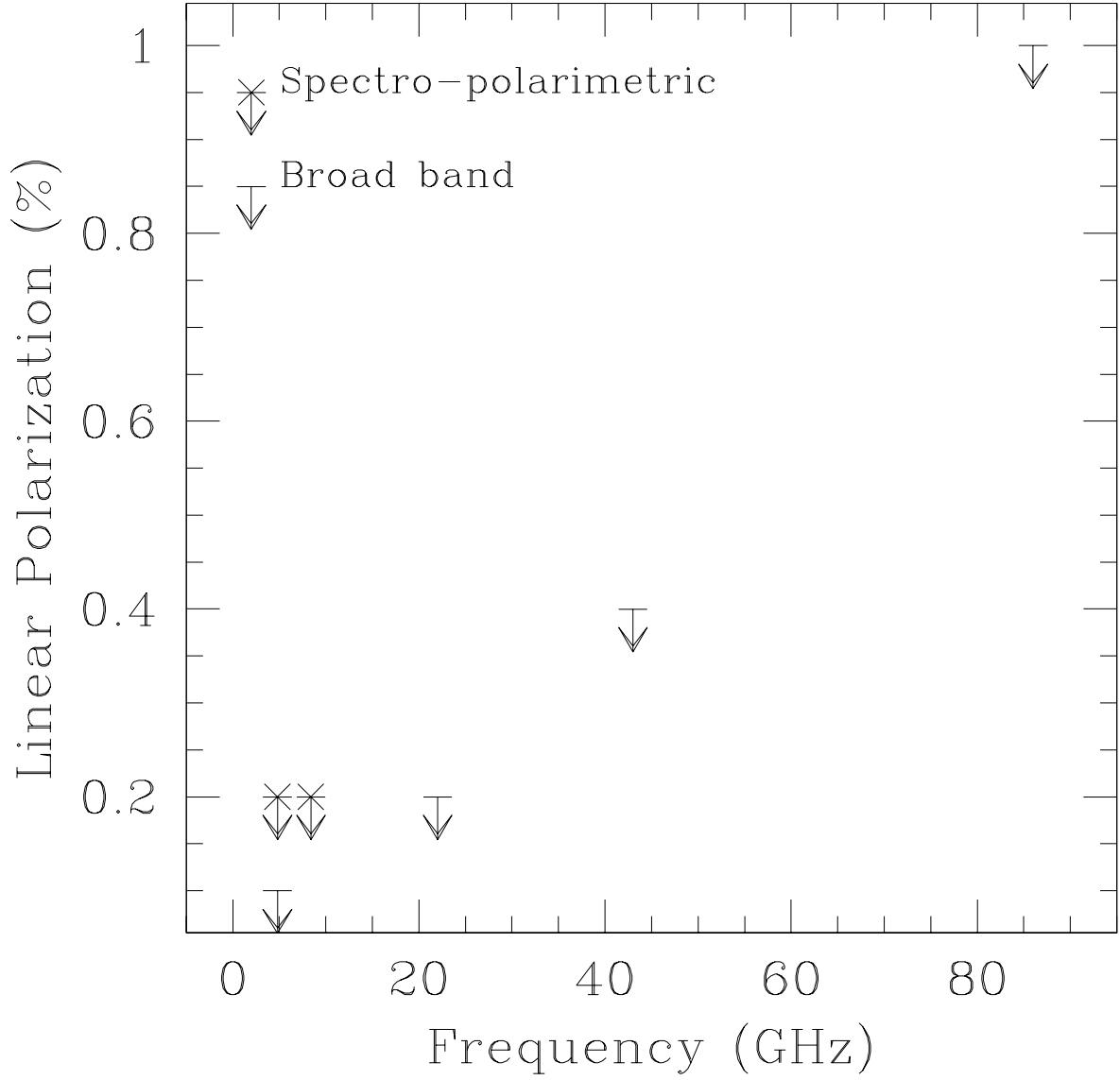,width=\textwidth}}
\caption{Upper limits to the linear polarization of
Sgr A*.  Broad band observations are indicated with an arrow.  
Spectro-polarimetric observations are indicated with an arrow and a
cross.}
\end{figure}

\newpage

\begin{deluxetable}{llrrrrrr}
\footnotesize
\tablecaption{Polarized and Total Flux from VLA Continuum Observations 
at 22 GHz and  43 GHz\label{tab:vlapol}}
\tablehead{
\colhead{Source}  &  \colhead{IF} & \colhead{$I$} & 
\colhead{$P$} & \colhead{$P_{lim}$} & \colhead{$p$} &\colhead{$p_{lim}$} &  \colhead{$\chi$} \\
                  &               & \colhead{(Jy)} & \colhead{(mJy)} & \colhead{(mJy)}
&   \colhead{(\%)} &   \colhead{(\%)} & \colhead{(deg)} \\
}
\startdata
\multicolumn{8}{c}{22 GHz} \\
\hline
B1730-130 & 1 & $11.342 \pm 0.072$ & $301.0 \pm 2.3 $ & $ 21.5$ 
&  $2.65 \pm 0.02 $ & $ 0.19$ & \dots  \\
	& 2 & $11.748 \pm 0.064$ & $281.4 \pm 2.4 $ & $ 26.7$  & $2.40 
\pm 0.02 $ & $ 0.23$ & \dots \\
B1741-312 & 1 & $0.663 \pm 0.001$  & $29.8 
\pm 1.0 $ & $ 2.9$ & $4.49 \pm 0.15 $ & $ 0.44$  & $-35.9 \pm 1.3$ \\
         & 2 &  $0.668 \pm 0.001$ & $28.2 
\pm 0.5 $ & $ 3.0$ & $4.22 \pm 0.08 $ & $ 0.45$ & $-38.9 \pm 0.7$ \\
Sgr A* & 1 & $1.053 \pm 0.001$ &  $2.0 \pm 0.5 $ & $ 2.1$
& $0.20 \pm 0.05 $ & $ 0.20$ & $-34.4 \pm 10.1$ \\
       & 2 & $1.061 \pm 0.001$ & $1.7 \pm 0.5 $ & $ 2.2$
& $0.16 \pm 0.05 $ & $ 0.21$ & $-12.3 \pm 11.9$ \\
\hline
\multicolumn{8}{c}{43 GHz} \\
\hline
B1730-130 & 1 & $11.512 \pm 0.004$ & $227.0 \pm 3.2 $ & $ 9.5$ & 
$1.96 \pm 0.03 $ & $ 0.08$ & \dots \\
         & 2 & $11.482 \pm 0.004$ & $219.5 \pm 4.5 $ & $ 11.3$ & 
$1.91\pm 0.04 $ & $ 0.10$ & \dots \\
B1741-312 & 1 & $0.476 \pm 0.002$ & $28.6 
\pm 1.2 $ & $ 3.5$ & $6.01 \pm 0.25 $ & $ 0.74$ & $9.4 \pm 1.7$ \\
         & 2 & $0.479 \pm 0.002$ & $29.3 
\pm 1.5 $ & $ 3.1$ & $6.12 \pm 0.31 $ & $ 0.65$ & $7.0 \pm 2.1$ \\
B1921-293 & 1 & $14.154 \pm 0.032$ & $115.6 
\pm 12.4 $ & $ 12.2$  & $0.82 \pm 0.09 $ & $ 0.09$ & $10.2 \pm 4.3$ \\
         & 2 & $14.161 \pm 0.032$ & $89.6  
\pm 15.3 $ & $ 17.9$ & $0.63 \pm 0.11 $ & $ 0.13$ & $8.2 \pm 6.9$ \\
Sgr A*   & 1 & $1.074 \pm 0.001$ & $3.4 \pm 0.9 $ & $ 2.6$
 & $0.32 \pm 0.08 $ & $ 0.24$ &  $-29.3 \pm 10.7$ \\
         & 2 & $1.073 \pm 0.001$ & $3.7 \pm 0.8 $ & $ 2.5$
& $0.34 \pm 0.08 $ & $ 0.23$ & $91.7 \pm 8.8$ \\
\enddata
\end{deluxetable}

\begin{deluxetable}{llrrrrrr}
\footnotesize
\tablecaption{Polarized and Total Flux from BIMA Continuum Observations 
at 86 GHz \label{tab:bimapol}}
\tablehead{
\colhead{Source}  &  \colhead{IF} & \colhead{$I$} &
\colhead{$P$} & \colhead{$P_{lim}$} & \colhead{$p$} &\colhead{$p_{lim}$} &  \colhead{$\chi$} \\
                  &               & \colhead{(Jy)} & \colhead{(mJy)} & \colhead{(mJy)} 
&   \colhead{(\%)} &   \colhead{(\%)} & \colhead{(deg)} \\
}
\startdata
\multicolumn{8}{c}{10 March 1998} \\
\hline
   3C273 & 1 & $25.250 \pm  0.340 $ & 
$742.8 \pm  13.2$ &39.3 & $ 2.94 \pm  0.05$ & 0.15 &  $  -2.1 \pm    0.5$ \\
 3C454.3 & 1 & $ 5.564 \pm  0.095 $ & 
$ 47.8 \pm   6.4$ &28.3 & $ 0.86 \pm  0.11$ & 0.51 & $  40.9 \pm    3.8$ \\
 Sgr A*  & 1 & $ 1.715 \pm  0.053 $ & 
$17.3 \pm 3.2$ & 23.1 & $1.01 \pm 0.18$ & 1.35 & $-32.6 \pm 7.5$ \\
\hline
\multicolumn{8}{c}{14 March 1998} \\
\hline
   3C273 & 1 & $23.180 \pm  0.368 $ &
$726.3 \pm   2.7$ & 44.0 & $ 3.13 \pm  0.01$ & 0.19 & $  -3.5 \pm    0.1$ \\
 3C454.3 & 1 & $ 4.816 \pm  0.105 $ &
$ 39.1 \pm  13.8$ & 37.8 & $ 0.81 \pm  0.29$ & 0.78 & $ 50.5 \pm   10.1$ \\
B1730-130 & 1 & $ 2.824 \pm  0.020 $ & 
$ 71.3 \pm   1.2$ & 67.4 & $ 2.53 \pm  0.04$ & 2.39 & $ 50.8 \pm    0.5$ \\
Sgr A*   & 1 & $ 1.352 \pm  0.065 $ & 
$9.5 \pm 7.0$ & 35.6 & $0.70 \pm 0.52$ & 2.63 & $-12.1 \pm 29.9$ \\
\hline
\multicolumn{8}{c}{19 December 1998} \\
\hline
   3C273 & 1 & $18.400 \pm  0.045 $ &
 $1166.1 \pm   4.7$ & 94.1 & $ 6.34 \pm  0.03$ & 0.51 & $ -38.5 \pm    0.1$ \\
         & 2 & $18.250 \pm  0.083 $ & 
 $1136.2 \pm   7.0$ & 75.3 & $ 6.23 \pm  0.04$ & 0.41 & $ -38.5 \pm    0.2$ \\
 3C454.3 & 1 & $ 5.828 \pm  0.119 $ &
$291.7 \pm  10.4$ & 34.3 & $ 5.00 \pm  0.18$ & 0.59 & $-85.2 \pm    1.0$ \\
         & 2 & $ 5.646 \pm  0.069 $ &
$338.7 \pm  19.2$ & 30.4 & $ 6.00 \pm  0.34$ & 0.54 & $-84.8 \pm    1.6$ \\
B1730-130 & 1 & $ 2.841 \pm  0.001 $ & 
$ 77.5 \pm  21.1$ & 33.8 & $ 2.73 \pm  0.74$ & 1.19 & $  10.8 \pm    7.8$ \\
         & 2 & $ 2.754 \pm  0.002 $ &
$ 85.5 \pm   7.7$ & 34.0 & $ 3.11 \pm  0.28$ & 1.23 & $  14.9 \pm    2.6$ \\
  Sgr A* & 1 & $ 2.374 \pm  0.016 $ & 
$22.8 \pm 5.7 $ & $ 20.9$ & $0.96 \pm 0.24 $ & $ 0.88$ & $26.5 \pm 10.1$ \\
         & 2 & $ 2.422 \pm  0.018 $ &
$22.9 \pm 3.6 $ & $ 19.6$ & $0.95 \pm 0.15 $ & $ 0.81$ & $-39.1 \pm 6.4$ \\
\enddata
\end{deluxetable}

\end{document}